\documentclass[aps,prl,amsmath,twocolumn]{revtex4}
\usepackage{bm}
\usepackage{graphicx}
\begin{document}
\draft
\title{Qubit Rotation by STIRAP}

\author{Z. Kis$^{(1, 2)}$ and F. Renzoni$^{(3)}$}

\address{\ $^{(1)}$ Department of Physics, Royal Institute of Technology (KTH),
                    Roslagstullsbacken 21, SE-10691 Stockholm, Sweden} 

\address{\ $^{(2)}$ Research Institute for
Solid State Physics and Optics, H-1525 Budapest, P.O. Box 49, Hungary
}

\address{\ $^{(3)}$ Laboratoire Kastler Brossel, D\'epartement de Physique de 
l'Ecole Normale Sup\'erieure, 24 rue Lhomond, 75231, Paris Cedex 05, France}

\date{\today{}}

\begin{abstract}   
  We introduce a novel procedure for qubit rotation, alternative to the
  commonly used method of Rabi oscillations of controlled pulse area. It is
  based on the technique of Stimulated Raman Adiabatic Passage (STIRAP) and
  therefore it is robust against fluctuations of experimental parameters.
  Furthermore, our work shows that it is in principle possible to perform
  quantum logic operations via stimulated Raman adiabatic passage. This opens
  up the search for a completely new class of schemes to implement logic
  gates.
\end{abstract}


\maketitle

The growing interest in quantum computation stimulates the search for schemes
to prepare and manipulate quantum states \cite{Bouwmeester}.  The basic
operation on a single qubit is the rotation.  This is usually achieved via
coherent Rabi oscillations.  For a two-level optical transition Rabi
oscillations are simply produced by direct coupling using a laser field. If
the qubit is formed by two ground states instead, Rabi flopping is then
implemented by Raman transitions via a virtual excited state, and a
fundamental logic gate has been demonstrated using this configuration
\cite{raman}.  Fundamental quantum logic operations based on (conditional)
Rabi flopping have also been implemented in cavity QED experiments
\cite{haroche}.  To perform qubit rotation utilizing Rabi flopping, a complete
control of the pulse area is required. An imperfect control leads obviously to
errors in the computation procedure.

In this work, we introduce a novel procedure for qubit rotation. It is based
on the technique of Stimulated Raman Adiabatic Passage (STIRAP \cite{review})
and therefore it is robust against fluctuations of several 
experimental parameters such as the pulse area. Furthermore, our proposal
shows that STIRAP is not only a powerful tool to transfer population between
quantum states \cite{review2}, including here the creation of coherent
superpositions \cite{marte,una98,vitanov99,renzoni,una01a} and entangled
states \cite{una01b}, but can also play a role in the implementation of a
quantum logic operation.

Consider the generic linear superposition of two long-living atomic 
states $|1\rangle$ and $|2\rangle$ 
\begin{equation}
|i\rangle\! = \!\alpha |1\rangle + \beta |2\rangle~.
\label{qubit}
\end{equation}
The state rotated about a unit vector $\bm n$ through an angle $\zeta$ reads
\begin{equation}
|f\rangle\! =\! \hat R_{\bm n}(\zeta)|i\rangle~,
\label{rotated}
\end{equation}  
with $\hat R_{\bm n}(\zeta)$ being an element of the $SU_2$ group
defined as
\begin{equation}\label{rotop}
  \hat R_{\bm n}(\zeta)\!=\!\exp\left(-i\frac{\zeta}{2}\, 
  {\bm n}\cdot\hat{\bm \sigma}\right)\!=\!\cos\frac{\zeta}{2}-
  i{\bm n}\cdot\hat{\bm \sigma}\sin \frac{\zeta}{2}~,
\end{equation}
where ${\bm n}\!=\!(\sin\vartheta\cos\varphi, \sin\vartheta\sin\varphi,
\cos\vartheta )$, and $\hat{\bm\sigma}\!=\!(\sigma_x, \sigma_y, \sigma_z )$
are the Pauli's spin operators $\sigma_x\!=\!|1\rangle\langle
2|+|2\rangle\langle 1|$, $\sigma_y\!=\!i(|2\rangle\langle 1|-|1\rangle\langle
2|)$, $\sigma_z\!=\!|1\rangle\langle 1|-|2\rangle\langle 2|$. 

Mapping of the state (\ref{qubit}) into the state (\ref{rotated}) can be
achieved easily using a single STIRAP sequence, as described for example in
Ref.~\cite{renzoni}. However, in that scheme the Stokes and pump pulses which
link the different states must satisfy well defined relationships with the
coefficients of the initial and final superpositions. Therefore, that
procedure cannot be considered as a general rotation, because for a given
pulse sequence, not every state will be transformed in the same way.  As it
will be shown here, the use of multiple STIRAP sequences allows indeed true
rotations, with the axis and angle of rotation being uniquely defined
by the parameters of the laser pulses.

\begin{figure}
\includegraphics[width=5cm]{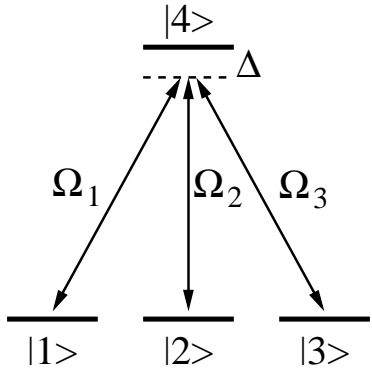}
\caption{Interaction scheme for the rotation of a qubit by STIRAP.
The three ground states $|1\rangle$, $|2\rangle$ and $|3\rangle$
are coupled to the excited state $|4\rangle$ by three different laser
fields. The qubit is defined by the ground states $|1\rangle$ and 
$|2\rangle$. State $|3\rangle$ is an auxiliary state occupied only in the 
intermediate phase of the rotation procedure.}
\label{scheme}
\end{figure}

In our scheme, shown in Fig. \ref{scheme}, three ground states
 $|1\rangle$, $|2\rangle$ and $|3\rangle$  are coupled via a
single excited state  $|4\rangle$  by different laser fields.
We assume that due to their polarizations and frequencies, each 
laser field drives only one transition. 
The ground states $|1\rangle$ and $|2\rangle$ define our qubit, while
the state $|3\rangle$ is an auxiliary state which will be occupied only 
in the intermediate phase of the rotation procedure.

The detunings of the three laser fields are the same, i.e. the system
is at multiphoton resonance which is a necessary condition for
STIRAP. The Schr\"odinger equation of this system reads
\begin{equation}\label{sch1}
  \frac{d}{dt}|\psi(t)\rangle\!=\!-\frac{i}{\hbar}{\hat
    H}(t)|\psi(t)\rangle,
\end{equation}
where $|\psi(t)\rangle$ denotes the state vector of the four-level
system and the Hamiltonian $\hat H(t)$ is given in the interaction
picture and in the rotating-wave approximation (RWA) as
\begin{equation}\label{ham}
  \hat H(t)\!=\!\hbar\Delta|4\rangle\langle 4|+\frac{\hbar}{2}
  \sum_{i=1}^3(\Omega_i(t)|i\rangle\langle 4|+h.c.)
\end{equation} 
Here $\Delta$ is the laser detuning.  The pulsed Rabi frequencies $\Omega_1
(t)$ and $\Omega_2 (t)$ are taken to have essentially the same envelopes:
$\Omega_1(t)\!=\!\Omega (t)\cos\chi$, $\Omega_2\!=\!\Omega
(t)\exp{(i\eta)}\sin\chi$, where $\chi$ and $\eta$ are fixed angles.  The
pulses 1,2 (i.e. the fields which yield the Rabi frequencies $\Omega_1(t)$ and
$\Omega_2(t)$) and 3 are delayed with respect to each other, however, for an
efficient STIRAP process they must have a significant overlap. In the
following $\Omega(t)$ will be taken as real.

Our procedure consists of two STIRAP processes. In the first one, the fields 1
and 2 are the pump fields, whereas the field 3 plays the role of the Stokes
field.  In this STIRAP process the field 3 has the same phase as that of
fields 1, 2. The pulses are applied in the counterintuitive order, i.e. the
Stokes pulse arrives before the pump ones. The transfer process can be
described as follows: The pump fields 1 and 2 define a dark (or noncoupled)
state $|NC\rangle $
\begin{equation}\label{nc}
  |NC\rangle\!=\!-\sin\chi |1\rangle + e^{i\eta}\cos\chi |2\rangle~,
\end{equation}
in the subspace spanned by the states $\{ |1\rangle ,|2\rangle \}$
\cite{ari}.  The orthogonal state (the coupled or bright state) 
$|C\rangle $ is
\begin{equation}\label{cc}
  |C\rangle\!=\!\cos\chi |1\rangle + e^{i\eta}\sin\chi |2\rangle~,
\end{equation}
from which the population can be transferred to state $|3\rangle$ if
all the three fields are on. By decomposing the initial superposition
$|i\rangle $ onto $|C\rangle $ and $|NC\rangle $,
\begin{equation}
|i\rangle = \langle NC|i\rangle |NC\rangle + \langle C|i\rangle |C\rangle 
\end{equation}
with
\begin{mathletters}
\begin{eqnarray}
  \langle NC|i\rangle &=&  -\alpha\sin\chi + \beta e^{-i\eta}\cos\chi \\
  \langle C|i\rangle &=& \alpha\cos\chi +\beta e^{-i\eta}\sin\chi
\end{eqnarray}
\label{coeff}
\end{mathletters}
we  find  that only the component along $|C\rangle $ will be
affected by the STIRAP process and eventually mapped to the target
state $|3\rangle $. Indeed, the Hamiltonian in Eq.~(\ref{ham})
can be rewritten in the form
\begin{equation}\label{hamnew}
  \hat H(t)\!=\!\hbar\Delta|4\rangle\langle 4|+\frac{\hbar}{2}
  (\Omega(t)|C\rangle\langle 4|+\Omega_3(t)|3\rangle\langle 4|+h.c.)~.
\end{equation}
Inserting Eq.~(\ref{hamnew}) into Eq.~(\ref{sch1}) one obtains a
Schr\"odinger equation which describes an ordinary STIRAP process in a
three-level system. The dark state $|\psi_D(t)\rangle$ of the Hamiltonian
in Eq.~(\ref{hamnew}) is given by
\begin{equation}\label{dark}
  |\psi_D(t)\rangle\!=\frac{1}{\sqrt{\Omega^2(t)+\Omega^2_3(t)}}
  (\Omega_3(t) |C\rangle - \Omega(t) |3\rangle)~.
\end{equation}
Using the same argument as in the case of STIRAP, it can be easily
shown that in the adiabatic limit the system is left, after the first
pulse sequence, in the superposition of the three ground states
\begin{equation}
  |\psi\rangle \!= \langle NC|i\rangle |NC\rangle
  - \langle C|i\rangle |3\rangle~, 
  \label{int}
\end{equation}
without populating the excited state during the evolution. 
We recognize in Eq. (\ref{int}) that the component of $|i\rangle $ along
the noncoupled state $|NC\rangle $ is untouched, whereas the orthogonal
bright component is transferred to the target state $|3\rangle $.

The second step of the rotation procedure is a reverse STIRAP process which
maps the state $|3\rangle$ back to the qubit subspace.  In this STIRAP process
the phase of field 3 is in general different from that of fields 1, 2.
The pulses are applied in the reverse order with respect to the first step
of the rotation procedure: the pulses 1 and 2 (which play now the role of the
Stokes pulses) arrive before the pulse 3 (the pump). The fields 1 and 2 have
the same Rabi frequencies as in the first step of the procedure. In this way
the state $|\psi\rangle$ prepared by the first STIRAP process,
Eq.~(\ref{int}), is initially a dark state for the three laser pulses because:
(a) the state $|3\rangle$ is initially not coupled (counterintuitive pulse
order); (b) the state $|\psi\rangle$ has no component along $|C\rangle$ and
therefore it is decoupled from the fields 1,2. The darkness of $|\psi\rangle $
allows the implementation of the second STIRAP process although all the ground
states are initially populated. In this process the state $|3\rangle $ is
transferred back to the qubit subspace. More precisely, it will be mapped on
the coupled state $|C\rangle$ because the state $|NC\rangle $ is a decoupled
state also in this second STIRAP process. The component of the initial qubit
$|i\rangle $ along the coupled state $|C\rangle $ and the new component along
the same state obtained by mapping back the state $|3\rangle $ will differ by
a phase.  This phase is determined by the phase difference $\delta$ between
the relative phase of the field 3 with respect to the fields 1,2 in the two
STIRAP processes.  Clearly, for identical phases the system certainly goes
back to the initial state $|i\rangle $.
In the general case a similar calculation which yielded Eq.~(\ref{int}) shows
that in the adiabatic limit the component of $|\psi\rangle$ along $|3\rangle $
is mapped back onto the subspace $\{ |1\rangle , |2\rangle \}$ according to
\begin{eqnarray}
  \langle 3|\psi\rangle|3\rangle &\to &   
  e^{-i\delta}\langle 3|\psi\rangle|C\rangle ~,
\end{eqnarray} 
so that the final state is
\begin{equation}
  |\psi_f\rangle \!=\! \langle NC|i\rangle |NC\rangle
  +e^{-i\delta}\langle C|i\rangle |C\rangle~.
\end{equation}
By substituting the expressions (\ref{coeff}) for the coefficients and
taking into account Eqs.~(\ref{rotated}) and (\ref{rotop}), we obtain
\begin{eqnarray}\label{psif}
  |\psi_f\rangle \!=\!e^{-i\delta/2} \hat R_{\bm n}(\delta)|i\rangle~,
\end{eqnarray}
where ${\bm n}\!=\!(\sin2\chi\cos\eta, \sin2\chi\sin\eta, \cos2\chi )$. Apart
from a global phase $-\delta/2$ the states $|i\rangle$ and $|\psi_f\rangle$
are connected through the rotation $\hat R_{\bm n}(\delta)$. If the qubit
is isolated, then this global phase is unimportant. If the qubit is part of a
larger system, e.g. there are several qubits which form a quantum computer,
then the global phase is clearly relevant, however, it may be incorporated
into the algorithm being implemented on the quantum computer.

The geometric interpretation of the rotation procedure described above is
straightforward: the axis of rotation ${\bm n}$ is defined by the polar
angle $2\chi$ and the azimutal angle $\eta$, which characterize the relative 
amplitude and phase of the laser pulses acting on the states $|1\rangle$ and 
$|2\rangle$, respectively. In the Hilbert space the state vector $|NC\rangle$, 
Eq. (\ref{nc}), is parallel with this axis, therefore, it is unaffected by 
the rotation. The state vector $|C\rangle$, Eq. (\ref{cc}), is perpendicular 
to the rotation axis, and it is rotated through an angle $\delta$ about 
this axis.

Our rotation procedure is sensitive to the relative phase and relative
amplitude of the pulses 1 and 2. Moreover, the method is also sensitive to the
phase $\delta$, the phase difference of the pulse 3 in the first and second
STIRAP processes. However, it is robust against the fluctuation of the pulse
shapes and pulse areas.  

\begin{figure}
\includegraphics[width=17cm]{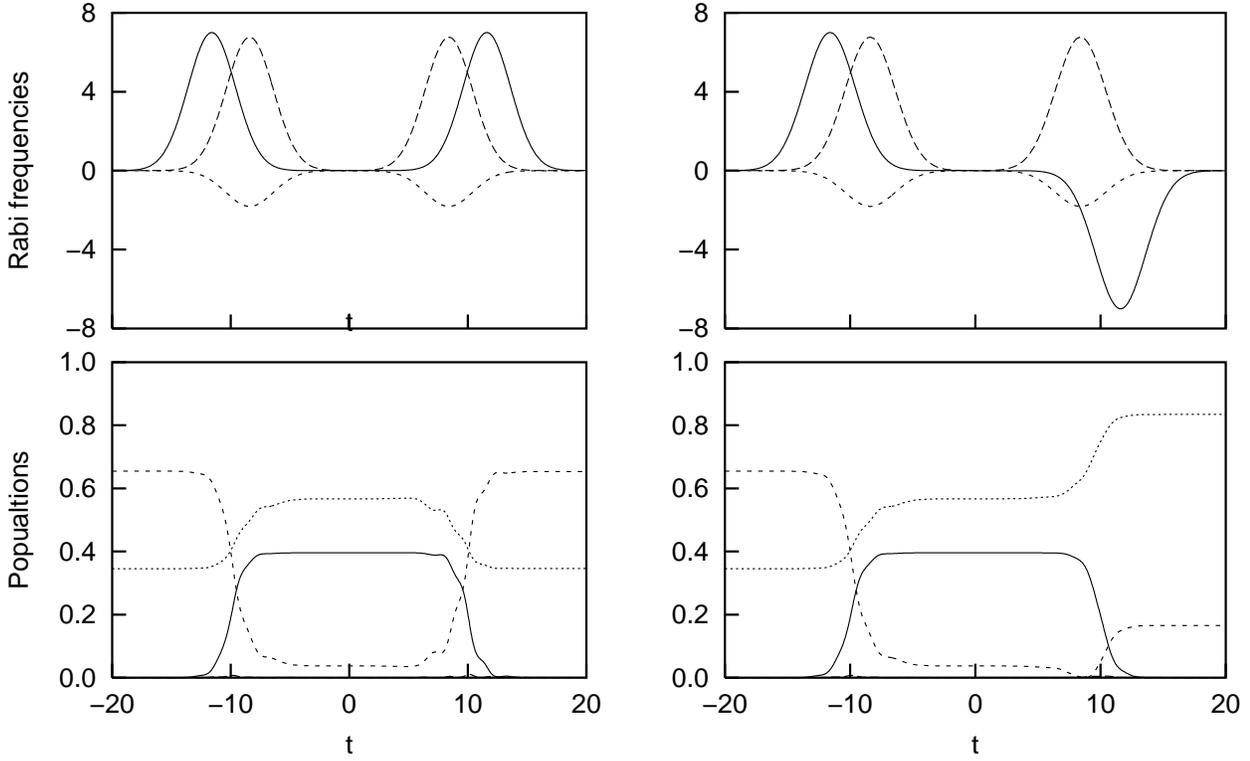}
\caption{
  Time evolution of the ground state populations and pulse sequences for two
  different STIRAP procedures. In the procedure shown in the left column, the
  second STIRAP process is precisely the reverse of the first one (top - left)
  and the system returns back to its original state (bottom - left). In the
  procedure shown in the right column, in the second STIRAP process the phase
  of the pulse 3 is shifted by $\delta\!=\!\pi$ with respect to the the pulses
  1 and 2 (top-right). In this case the system does not return back to its
  original state, and the qubit is rotated through an angle $\pi$ about a unit
  vector characterized by the polar angle $2\chi$ and the azimutal angle 
  $\eta$. Parameters of the calculation are: $\alpha\!=\! \cos\pi/5$,
  $\beta\!=\!\sin\pi/5$, $\Delta\!=\!0$, $\chi\!=\!-\pi/12$, and $\eta\!=\!0$.
  The pulses have Gaussian shape $\exp[ -(t\pm T/2 \pm t_0)^2/2\tau^2 ]$, with
  $\tau\!=\!2$, $t_0\!=\!1.6$, $T\!=\!20$. The time and the Rabi frequencies 
  are measured in arbitrary units.  }
\label{numerics}
\end{figure}

Our analysis is supported by numerical calculations.  In these calculations,
the envelope of the different pulsed Rabi frequencies has been taken as
Gaussian, although any sufficiently smooth pulse shape is suitable. The width
of the pulses and their overlaps have been chosen to satisfy well the usual
STIRAP conditions \cite{review}.  For different choices of the initial state
of the qubit (\ref{qubit}) we have solved numerically the Schr\"odinger
equation (\ref{sch1}) and determined the time dependent state of the system
during the rotation procedure composed of two STIRAP processes. An example of
our calculations is shown in Fig.~\ref{numerics} where the populations of the
three ground states are shown together with the pulse sequence which effects
the qubit rotation. The time evolution in the case of identical phases for the
two STIRAP processes is also reported for comparison. In order to analyze
quantitatively the quality of our rotation procedure, we have calculated the
fidelity $F=|\langle\psi_f|\psi_{\infty}\rangle |^2$, where
$|\psi_{\infty}\rangle $ is the numerically calculated state after the
rotation procedure and $|\psi_f\rangle $ the state of the rotated qubit. We
have verified that the fidelity is unity within numerical error.  We have also
verified that under usual STIRAP conditions the excited state occupation is
negligeable throughout the rotation procedure.

In conclusion we have shown that it is possible to perform qubit rotations by
stimulated Raman adiabatic passage, and correspondingly, proposed a rotation
procedure. The procedure is composed of two STIRAP processes. The first step
consists in projecting the original qubit on two perpendicular states: the
dark and the bright states, as defined by the pump pulses. The component of
the qubit along the bright state is then transferred to an auxiliary state. In
the second STIRAP process, the auxiliary state is mapped back onto the bright
state, with a phase shift $\delta$ with respect to the initial component of
the qubit along this state.  The resulting state corresponds indeed to a
rotation of the qubit, with the axis and angle of rotation determined 
uniquely by the parameters of the laser fields.

Furthermore, our work shows that it is in principle possible to perform 
quantum logic operations via stimulated Raman adiabatic passage. This 
opens up the search for new schemes for logic gates based on STIRAP.

We are indebted to Stig Stenholm for useful discussions and a careful reading
of the manuscript. F.R. is also grateful for his hospitality at the Royal
Institue of Technology (KTH) of Stockholm, where the main part of this work
has been done. Z. Kis acknowledges the financial support of the European Union
Research and Training Network COCOMO, contract HPRN-CT-1999-00129, the J\'anos
Bolyai program of the Hungarian Academy of Sciences, and the National
Scientific research Fund (OTKA) of Hungary under Contract No. T034484.

\newpage

\end{document}